\documentclass[twocolumn,prl,floatfix,tighten,bibtex,letterpaper]{revtex4}%
\usepackage{graphicx}
\usepackage{bm}
\usepackage{amssymb}
\usepackage{color}
\usepackage{epsfig}
\usepackage{subfigure}
\usepackage{amsmath}
\usepackage{amsfonts}%
\providecommand{\U}[1]{\protect\rule{.1in}{.1in}}

\begin{document}
\title{Magneto-Josephson effects in junctions with Majorana bound states}
\author{Liang Jiang$^{1}$, David Pekker$^{1}$, Jason Alicea$^{2}$, Gil Refael$^{1,5}$,
Yuval Oreg$^{3}$, Arne Brataas$^{4}$, and Felix von Oppen$^{5}$}
\affiliation{$^{1}$Department of Physics, California Institute of Technology, Pasadena,
California 91125, USA}
\affiliation{$^{2}$ Department of Physics and Astronomy, University of California, Irvine,
CA 92697}
\affiliation{$^{3}$Department of Condensed Matter Physics, Weizmann Institute of Science,
Rehovot, 76100, Israel }
\affiliation{$^{4}$ Department of Physics, Norwegian University of Science and Technology,
N-7491 Trondheim, Norway}
\affiliation{$^{5}$Dahlem Center for Complex Quantum Systems and Fachbereich Physik, Freie
Universit\"{a}t Berlin, 14195 Berlin, Germany}
\date{\today}

\begin{abstract}
We investigate 1D quantum systems that support Majorana bound states at
interfaces between topologically distinct regions. In particular, we show that
there exists a duality between particle-hole and spin degrees of freedom in
certain spin-orbit-coupled 1D platforms such as topological insulator edges.
This duality results in a spin analogue of previously explored `fractional
Josephson effects'---that is, the \emph{spin current} flowing across a
magnetic junction exhibits $4\pi$ periodicity in the relative magnetic field
angle across the junction. Furthermore, the interplay between the
particle-hole and spin degrees of freedom results in unconventional
magneto-Josephson effects, such that the Josephson current is a function of
the magnetic field orientation with periodicity $4\pi$.

\end{abstract}
\maketitle


The possibility of observing Majorana zero-modes in condensed matter has
captured a great deal of attention in recent years. Much effort in this
pursuit presently focuses on spin-orbit-coupled 1D wires, which are closely
related to edges of 2D topological insulators (TIs). In either setting
Majorana modes are predicted to localize through the competition between
superconducting proximity effects and Zeeman splitting
\cite{FuL09c,Lutchyn10,Oreg10,JPAROV11,Beenakker11,Alicea12}. Remarkably,
zero-bias conductance anomalies \cite{Sengupta01, Bolech07,Law09, Flensberg10,
Fidkowski12} possibly originating from Majorana modes have even been measured
\cite{Kouwenhoven12,Das12} very recently in quantum wires. Numerous other
fascinating phenomena tied to Majorana fermions have also been explored,
including non-Abelian statistics \cite{Read00,Ivanov01,Alicea11}, electron
teleporation \cite{FuL10}, and exotic Josephson effects
\cite{Kitaev01,FuL09c,JPAROV11}.

\begin{figure}[ptb]
\centering
\includegraphics[width=8.8cm]{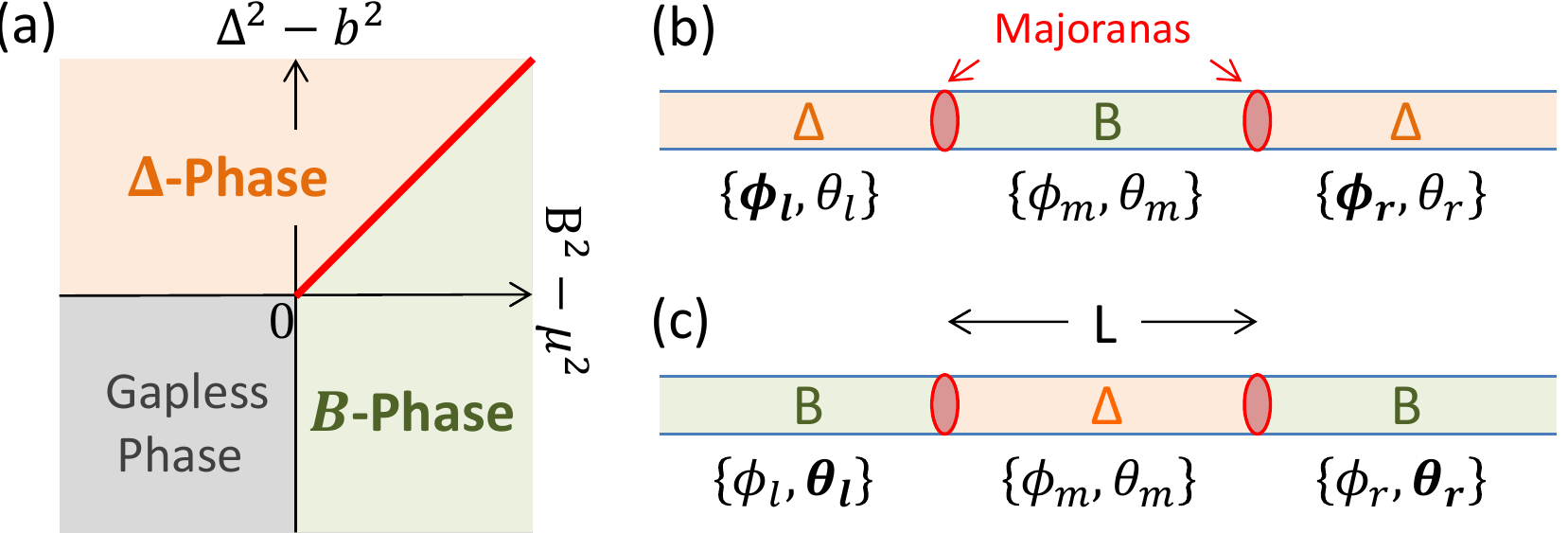}\caption[fig:PhaseDiagram]%
{(a) Phase diagram for 1D system: gapless-phase ($B^{2}\leq\mu^{2}$ and
$\Delta^{2}\leq b^{2})$, $\mathrm{\Delta}$-phase ($\Delta^{2}-b^{2}%
>\max\left[  B^{2}-\mu^{2},0\right]  $), and \textrm{B}-phase ($B^{2}-\mu
^{2}>\max\left[  \Delta^{2}-b^{2},0\right]  $). Both $\mathrm{\Delta}$-phase
and \textrm{B}-phase are gapped. (b) The $\mathrm{\Delta}$-\textrm{B}%
-$\mathrm{\Delta}$ junction supports Majorana bound states at the domain walls
\cite{JPAROV11}. (c) The dual configuration of \textrm{B}-$\mathrm{\Delta}%
$-\textrm{B} junction that also supports Majoranas.}%
\label{fig:PhaseDiagram}%
\end{figure}

Particularly interesting to us here are the Majorana-related Josephson effects
in quantum wires and TI edges. Consider two Majorana modes hybridized across a
Josephson junction formed by topological superconducting regions separated by
a narrow barrier as shown in Fig.~\ref{fig:PhaseDiagram}(b). The energy
splitting of these Majoranas depends periodically on \textit{half} the phase
difference between the right and left superconductors, $(\phi_{r}-\phi_{l}%
)/2$, giving rise to a Josephson current with $4\pi$ periodicity in $\phi
_{r}-\phi_{l}$ \cite{Kitaev01,FuL09c}. If, in addition, a third superconductor
contacts the middle domain, a difference between its phase and the
\emph{average} phase $(\phi_{r}+\phi_{l})/2$ induces a non-local three-leg
\textquotedblleft zipper\textquotedblright\ Josephson current that divides
equally between the two leads and is also $4\pi$ periodic in $\phi_{r}$ and
$\phi_{l}$ \cite{JPAROV11}. These `fractional Josephson effects' provide
smoking-gun signatures of Majorana modes.

Our claim is that physical quantities of Majorana junctions in wires and TI
edges can also possess $4\pi$-periodic dependence on the \textit{orientations}
of Zeeman fields applied in the plane normal to the spin orbit direction.
Notably, in some domain configurations the Majorana-mediated Josephson current
\emph{reverses sign} after a full $2\pi$ rotation of the magnetic field
orientation on one side of the junction. Only an additional $2\pi$ rotation
restores the currents to their original direction. Thus the mixing between the
particle-hole and spin degrees of freedom leads to an \emph{unconventional
magneto-Josephson effect} through the coupling of Majoranas.

Additionally, the Majorana modes produce a `spin Josephson current' between
the magnets providing the Zeeman energy, which could also be $4\pi$ periodic
in the field orientations. Define $\theta_{s}$ as the angle between the wire
and the Zeeman field at domain $s$. Spin Josephson currents, $j^{S}$, are
equivalent to torques (driven partly by the Majoranas) that the wire domains
apply on the external magnets \footnote{Technically, interpreting Eq. (1) as a
spin current is valid when one employs topological insulators with globally
conserved $S^{Z}$.}. Therefore, they are given by the derivative of the
system's energy with respect to the magnetic field orientations $\theta$:
\begin{equation}
j^{S}=\frac{\partial\left\langle {\mathcal{H}}\right\rangle }{\partial\theta
}.\label{eq:SpinCurrent}%
\end{equation}
with $\mathcal{H}$ being the system's Hamiltonian. In the case of TI edges,
the spin currents arise as the exact duals of Josephson currents, and the
orientation of the B-field is the exact dual to the superconducting phase
(indeed, the Josephson current is given by $j^{Q}=\frac{2e}{\hbar}%
\frac{\partial\left\langle {\mathcal{H}}\right\rangle }{\partial\phi}%
$)\footnote{In fact, a similar duality can be constructed for a tight-binding
description of a spin-orbit coupled quantum wire.}. We emphasize that the
$4\pi$ periodicity prevails as long as the parity of the Majorana state
remains constant during the measurement, or changes at a slower rate than the
winding of the superconducting phase and magnetic orientations.


Let us focus first on the analysis of the $4\pi$-periodic orientation
dependence in TI edges, before commenting on spin-orbit-coupled wires which
obey qualitatively similar rules. The Hamiltonian, including s-wave pairing
and Zeeman fields in both the transverse and parallel directions relative to
the spin-orbit direction, reads
\begin{align}
{\mathcal{H}}=v\hat{p}\tau^{z}\sigma^{z} &  -\mu\tau^{z}+\Delta\left(
\cos\phi~\tau^{x}-\sin\phi~\tau^{y}\right)  \nonumber\\
&  -b\sigma^{z}+B\left(  \cos\theta~\sigma^{x}-\sin\theta~\sigma^{y}\right)
.\label{eq:DefH}%
\end{align}
Here we have employed the Nambu spinor basis $\Psi^{T}=(\psi_{\uparrow}%
,\psi_{\downarrow},\psi_{\downarrow}^{\dagger},-\psi_{\uparrow}^{\dagger})$
and introduced Pauli matrices $\sigma^{a}$ and $\tau^{a}$ that act in the spin
and particle-hole sectors, respectively. The edge-state velocity is given by
$v$, $\hat{p}$ is the momentum, and the $\sigma^{z}$-direction represents the
spin-orbit-coupling axis. We allow the chemical potential $\mu$,
superconducting pairing $\Delta e^{i\phi}$, longitudinal magnetic field
strength $b$, transverse magnetic field strength $B$, and the transverse-field
orientation angle $\theta$ to vary spatially.
Interestingly, Eq.\ (\ref{eq:DefH}) has a \emph{magnetism-superconductivity
duality}---the Hamiltonian takes the same form upon interchanging the magnetic
terms $\left\{  b,B,\theta,\sigma^{a}\right\}  $ with the superconducting
terms $\left\{  \mu,\Delta,\phi,\tau^{a}\right\}  $.
Below we deduce the physical consequences of this duality.

The Hamiltonian (\ref{eq:DefH}) supports three different phases determined by
the relative strength of $\left\{  \Delta,\mu,B,b\right\}  $. As
Fig.~\ref{fig:PhaseDiagram}(a) illustrates, we have (i) a topological
superconducting gapped phase (denoted henceforth as the $\mathrm{\Delta}%
$-phase) when $\Delta^{2}-b^{2}>\max\left[  B^{2}-\mu^{2},0\right]  $, (ii) a
topological magnetic gapped phase (denoted \textrm{B}-phase) when $B^{2}%
-\mu^{2}>\max\left[  \Delta^{2}-b^{2},0\right]  $, and (iii) a trivial gapless
state when $B^{2}\leq\mu^{2}$ and $\Delta^{2}\leq b^{2}$. Consistent with the
magnetic-superconducting duality, in the phase diagram of
Fig.~\ref{fig:PhaseDiagram}(a) the \textrm{B}- and $\mathrm{\Delta}$-phases
are symmetrically arranged with respect to the diagonal line that defines the
boundary between these two gapped states:%
\begin{equation}
\Delta^{2}+\mu^{2}=B^{2}+b^{2}.
\end{equation}
Majorana zero-modes bind to domain walls separating \textrm{B-} and
$\mathrm{\Delta}$-domains. For notational simplicity, below we will assume
that $\Delta>b>0$ and $B>\mu>0$, though more general results can be obtained
\cite{Appendix}. We will also focus on setups for which all domains experience
both superconductivity and a transverse Zeeman field.

In TI edges, the $4\pi$ periodic dependence on the magnetic field orientation
occurs when two Majoranas are nestled in a $\mathrm{B-\Delta-B}$ domain
sequence as in Fig.~\ref{fig:PhaseDiagram}(c). This is in contrast to the
previously studied unconventional Josephson effects
\cite{Kitaev01,FuL09c,JPAROV11}, which occur over a junction between two
$\mathrm{\Delta}$-domains bridged by a \textrm{B}-domain [see
Fig.~\ref{fig:PhaseDiagram}(b)]. The magneto-Josephson and spin-Josephson
effects of a TI edge follow from the detailed dependence of the Majorana
energy splitting, ${E_{\mathrm{Maj}}}$, on the field orientations and
superconducting phases in the $\mathrm{B-\Delta-B}$ edge domain structure of
Fig.~\ref{fig:PhaseDiagram}(c). In addition to an exact numerical calculation
of ${E_{\mathrm{Maj}}}$, we provide in \cite{Appendix} an analytical
variational approach that sheds light on the physics. In the latter approach
we assume that the Majorana wavefunctions are unmodified by their proximity to
each other, apart from being superposed to form a conventional low-lying
state. This leads to an energy splitting that is suppressed as a weighted sum
of two exponentials which control the decay of the Majorana wave functions in
the middle domain.

Our results for the Majorana couplings constitute one of the central results
of this paper.
The \emph{two} characteristic decay lengths as a function of field and pairing
are $\lambda_{1,2}=\frac{v}{|\sqrt{\Delta^{2}-b^{2}}\pm\sqrt{B^{2}-\mu^{2}}|}%
$. Quite generally, for the middle $\mathrm{\Delta}$-domain of length $L$, the
Majorana coupling energy is:%
\begin{align}
\frac{E_{\mathrm{Maj}}}{E_{0}\left[  \delta\phi_{l,r}\right]  }  &  \approx
e^{-\lambda_{m,1}L}\sin\frac{\delta\theta_{l}-\tilde{\mu}_{m}+\tilde{\mu}_{l}%
}{2}\sin\frac{\delta\theta_{r}+\tilde{\mu}_{m}-\tilde{\mu}_{r}}{2}\nonumber\\
&  -e^{-\lambda_{m,2}L}\sin\frac{\delta\theta_{l}+\tilde{\mu}_{m}+\tilde{\mu
}_{l}}{2}\sin\frac{\delta\theta_{r}-\tilde{\mu}_{m}-\tilde{\mu}_{r}}{2}.
\label{eq:Majo}%
\end{align}
Here we have defined $\delta\phi_{\ell,r}\equiv\phi_{\ell,r}-\phi_{m}$,
$\delta\theta_{\ell,r}\equiv\theta_{\ell,r}-\theta_{m}$, $\tilde{\mu}%
_{l/m/r}\equiv\cos^{-1}\frac{\mu_{l/m/r}}{B_{l/m/r}}$, $\tilde{b}%
_{l/m/r}\equiv\cos^{-1}\frac{b_{l/m/r}}{\Delta_{l/m/r}}$, along with a
characteristic energy%
\begin{equation}
E_{0}\left[  \delta\phi_{l,r}\right]  =\frac{\sin\tilde{b}_{m}}{\sin\tilde
{\mu}_{m}}\frac{1}{\sqrt{M_{l}\left[  \delta\phi_{l}\right]  M_{r}\left[
\delta\phi_{r}\right]  }}.
\end{equation}
The denominator of $E_{0}$ follows from%
\begin{align}
M_{s}  &  \left[  \delta\phi_{s}\right]  \approx\left.
\begin{array}
[c]{c}%
\frac{\left(  \Delta_{m}^{2}+\mu_{m}^{2}-b_{m}^{2}\right)  }{2\sqrt{\Delta
_{m}^{2}-b_{m}^{2}}\left(  \Delta_{m}^{2}+\mu_{m}^{2}-B_{m}^{2}-b_{m}%
^{2}\right)  }\\
\end{array}
\right. \\
&  \left.
\begin{array}
[c]{c}%
+\frac{\left(  B_{s}^{2}+b_{s}^{2}-\mu_{s}^{2}\right)  +\Delta_{s}\left[
\sqrt{B_{s}^{2}-\mu_{s}^{2}}\sin\left(  \tilde{b}_{m}\pm\delta\phi_{s}\right)
-b_{s}\cos\left(  \tilde{b}_{m}\pm\delta\phi_{s}\right)  \right]  }%
{2\sqrt{B_{s}^{2}-\mu_{s}^{2}}\left(  B_{s}^{2}+b_{s}^{2}-\Delta_{s}^{2}%
-\mu_{s}^{2}\right)  },\\
\end{array}
\right. \nonumber
\end{align}
with the choice of sign $\pm$ depending on $s=l$ or $r$. Note that $M_{s}$
exhibits the standard $2\pi$ periodicity in $\phi_{s}$, so that the more
exotic $4\pi$ periodicity follows exclusively from the trigonometric functions
in Eq.~(\ref{eq:Majo}).

\begin{figure}[ptb]
\centering
\includegraphics[width=8.5cm]{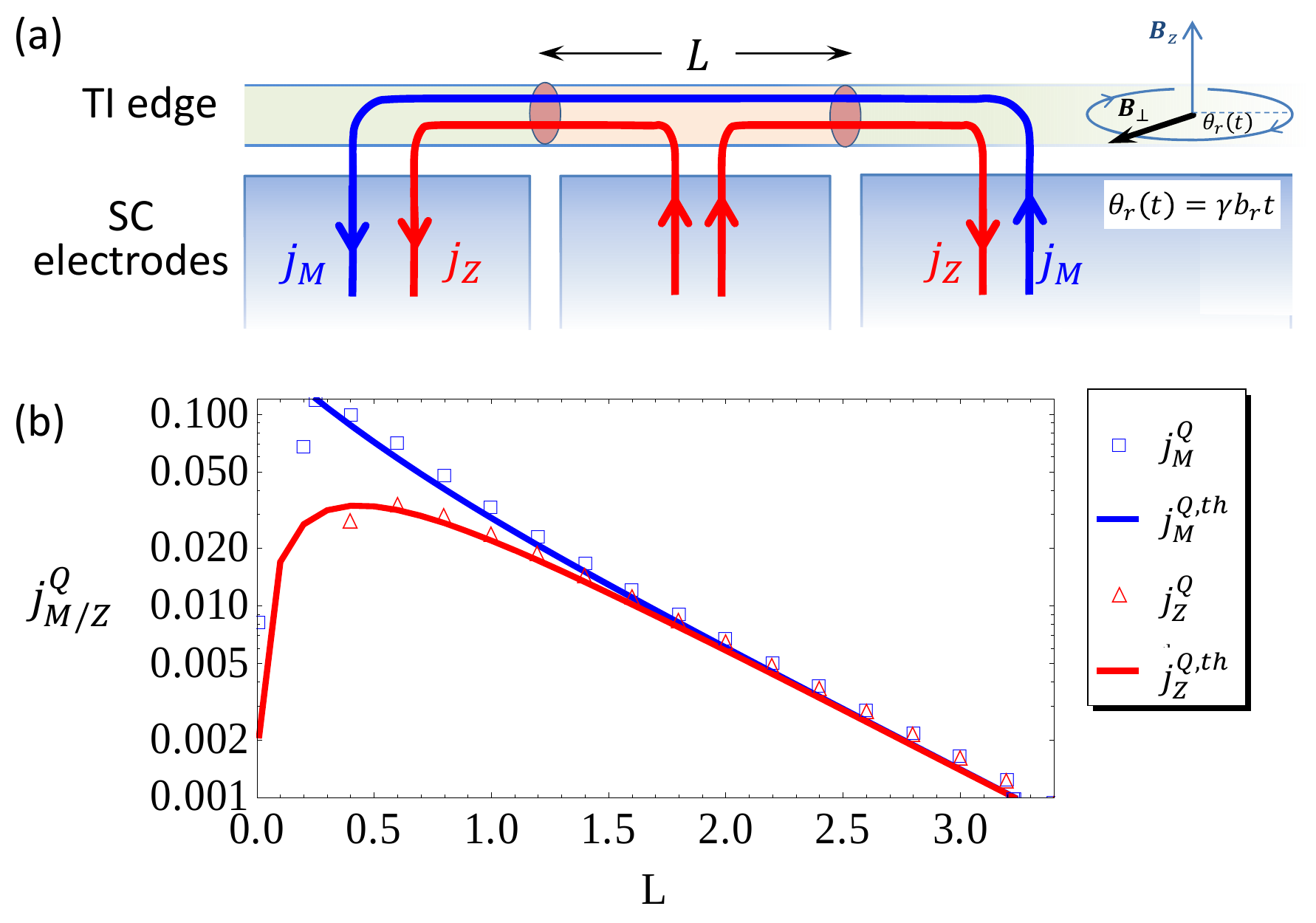}\caption[fig:JvsL]{ (a) The scheme
to measure unconventional magneto-Josephson effect. Josephson currents are
measured for the \textrm{B}-$\mathrm{\Delta}$-\textrm{B} junction. In the
right region, the transverse magnetic field winds at rate $\omega_{L}=\gamma
b_{r}$, which modulates the Josephson current at \emph{half} the frequency,
$\omega_{L}/2$. (b) Comparison between analytical expressions and numerical
results for $j_{M}$ and $j_{Z}$. The parameters are $\mu_{l/m/r}=0$,
$b_{l/m/r}=1/2$, $\Delta_{m}=2.5$, $\Delta_{l/r}=1$, $B_{l/r}=2$, $B_{m}=1$.
The superconducting angles are fixed $\phi_{l/r}=\pi/2$, $\phi_{m}=0$. }%
\label{fig:JvsL}%
\end{figure}

These general results allow us to quantitatively estimate the
magneto-Josephson effects described earlier, which can be measured in the
circuit sketched in Fig.~\ref{fig:JvsL}(a). For simplicity, we specialize to
the case of $\mu_{l/m/r}=0$, where the Majorana coupling energy reduces to
\begin{equation}
E_{\mathrm{Maj}}\approx\epsilon_{M}[\delta\phi_{\ell,r}]\cos\frac{\theta
_{l}-\theta_{r}}{2}+\epsilon_{Z}[\delta\phi_{\ell,r}]\cos\frac{\theta
_{l}+\theta_{r}-2\theta_{m}}{2}, \label{eq:Majo2}%
\end{equation}
with $\epsilon_{M/Z}[\delta\phi_{\ell,r}]=E_{0}[\delta\phi_{\ell,r}%
]\frac{e^{-L/\lambda_{m,2}}\pm e^{-L/\lambda_{m,1}}}{2}$.


The Majorana-related magneto-Josephson currents entering the $s=\ell/r$
electrode are $j_{s}^{Q}=\frac{2e}{\hbar}\frac{\partial\left\langle
{\mathcal{H}}\right\rangle }{\partial\phi_{s}}=p\frac{2e}{\hbar}%
\frac{E_{\mathrm{Maj}}}{\partial\phi_{s}}$, where $p=\pm1$ denotes the parity
of the hybridized Majoranas. The explicit form for the charge currents
(dropping the parity factor $p$) is:
\begin{equation}%
\begin{array}
[c]{c}%
j_{\ell/r}^{Q}\approx\pm j_{M}^{Q}\cos\frac{\theta_{l}-\theta_{r}}{2}%
+j_{Z}^{Q}\cos\frac{\theta_{l}+\theta_{r}-2\theta_{m}}{2},\\
\text{\textrm{with:}}\mathrm{\,\,\,}j_{M/Z}^{Q}=\frac{2e}{\hbar}\frac
{\partial\epsilon_{M/Z}}{\partial\phi_{\ell/r}}.
\end{array}
\label{eq:MEpred}%
\end{equation}
which constitutes a prediction for the unconventional magneto-Josephson
effect. The analytical expressions obtained above for $j_{M}^{Q}$ and
$j_{Z}^{Q}$ agree well with the numerical calculations for large $L$ as shown
in Fig.~\ref{fig:JvsL}(b). They confirm that for \textrm{B}-$\mathrm{\Delta}%
$-\textrm{B} junctions the Majorana coupling induces the charge current
$j_{l/r}^{Q}$ with $4\pi$ periodic dependence on $\theta_{l/r}$.


Similarly the spin Josephson currents, or torques on the magnets, in region
$s=\ell/r$ are $j_{s}^{S}=-\frac{\partial\left\langle {\mathcal{H}%
}\right\rangle }{\partial\theta_{s}}=p\frac{\partial E_{\mathrm{Maj}}%
}{\partial\theta_{s}}$ [Eq.~(\ref{eq:SpinCurrent})]. The angular momentum
transferred by these currents is in the direction parallel to the spin-orbit
axis, which in this case is the $z$-direction. The spin Josephson currents are
thus given by:
\begin{equation}%
\begin{array}
[c]{c}%
j_{l/r}^{S}=\pm j_{M}^{S}\sin\frac{\theta_{l}-\theta_{r}}{2}+j_{Z}^{S}%
\sin\frac{\theta_{l}+\theta_{r}-2\theta_{m}}{2},\\
\text{\textrm{with:}}\mathrm{\,\,\,}j_{M/Z}^{S}=\frac{\epsilon_{M/Z}}{2}.
\end{array}
\end{equation}
The $j_{M}^{S}$ spin current exchanges angular momentum between the right and
left magnets directly, while the $j_{Z}^{S}$ spin current originates in the
middle region and equally splits into the right and left regions,
$j_{m\rightarrow l}^{S}=j_{m\rightarrow r}^{S}=j_{Z}^{S}\sin\frac{\theta
_{l}+\theta_{r}-2\theta_{m}}{2}$. This term vanishes when there is no
transverse magnetic field in the middle domain, and represents the dual of the
zipper Josephson effect in the $\mathrm{\Delta}$-\textrm{B}-$\mathrm{\Delta}$
junction that splits charge current from the middle domain between the two
side domains \cite{JPAROV11}.

The origin of this exotic dependence of the Majorana-related currents can be
traced to the magnetic-superconducting duality in topological insulator edges
\cite{FuL09c,JPAROV11}. For a junction with three alternating domains, there
are two dual configurations: the $\mathrm{\Delta}$-\textrm{B}-$\mathrm{\Delta
}$ junction [Fig.~\ref{fig:PhaseDiagram}(b)] and the \textrm{B}%
-$\mathrm{\Delta}$-\textrm{B} junction [Fig.~\ref{fig:PhaseDiagram}(c)]. The
spin-Josephson effect in the \textrm{B}-$\mathrm{\Delta}$-\textrm{B} junction
is dual to the charge-Josephson effect in the $\mathrm{\Delta}$-\textrm{B}%
-$\mathrm{\Delta}$ junction \cite{FuL09c,Lutchyn10,Oreg10,JPAROV11}.
Similarly, the magneto-Josephson effect depending on the orientation angles in
the \textrm{B}-$\mathrm{\Delta}$-\textrm{B} junction has a dual spin-Josephson
effect depending on the superconducting angles in the $\mathrm{\Delta}%
$-\textrm{B}-$\mathrm{\Delta}$ junction.


Majorana junctions in spin-orbit coupled wires exhibit the same
magneto-Josephson and spin-Josephson effects as the TI edge. The wire's
Hamiltonian adds a kinetic energy piece to Eq.~(\ref{eq:DefH}), $\mathcal{H}%
_{k}=\frac{1}{2m}{\hat{p}}^{2}\tau^{z}$. This produces additional Fermi points
at `large' momenta $p_{F}\sim\pm2mv$ that are, however, nearly unaffected by
the magnetic field in the presence of pairing. Therefore the analysis above
for the TI edges still applies qualitatively. Thus, in a Majorana wire, $4\pi
$-periodic effects in both $\theta$ and $\phi$ appear in the $B-\Delta-B$
domain sequence \footnote{In contrast to the $\phi$ dependent $4\pi$ periodic
Josephson effect, it requires \textit{opposite} domain sequence in a TI edge
($\mathrm{\Delta}$-\textrm{B}-$\mathrm{\Delta}$) and in wires (\textrm{B}%
-$\mathrm{\Delta}$-\textrm{B}) \cite{JPAROV11}. This arises since the paired
large-momentum Fermi points form a p-wave superconductor. If the $p=0$
crossing forms another p-wave superconductor, together the two form a
topologically trivial phase.}. The quantitative analysis of the magneto-,
spin-, and charge-Josephson effects in wires as well as the role of Andreev
bound states will be analyzed elsewhere \cite{InPreparation}.



\begin{figure}[ptb]
\includegraphics[width=8cm]{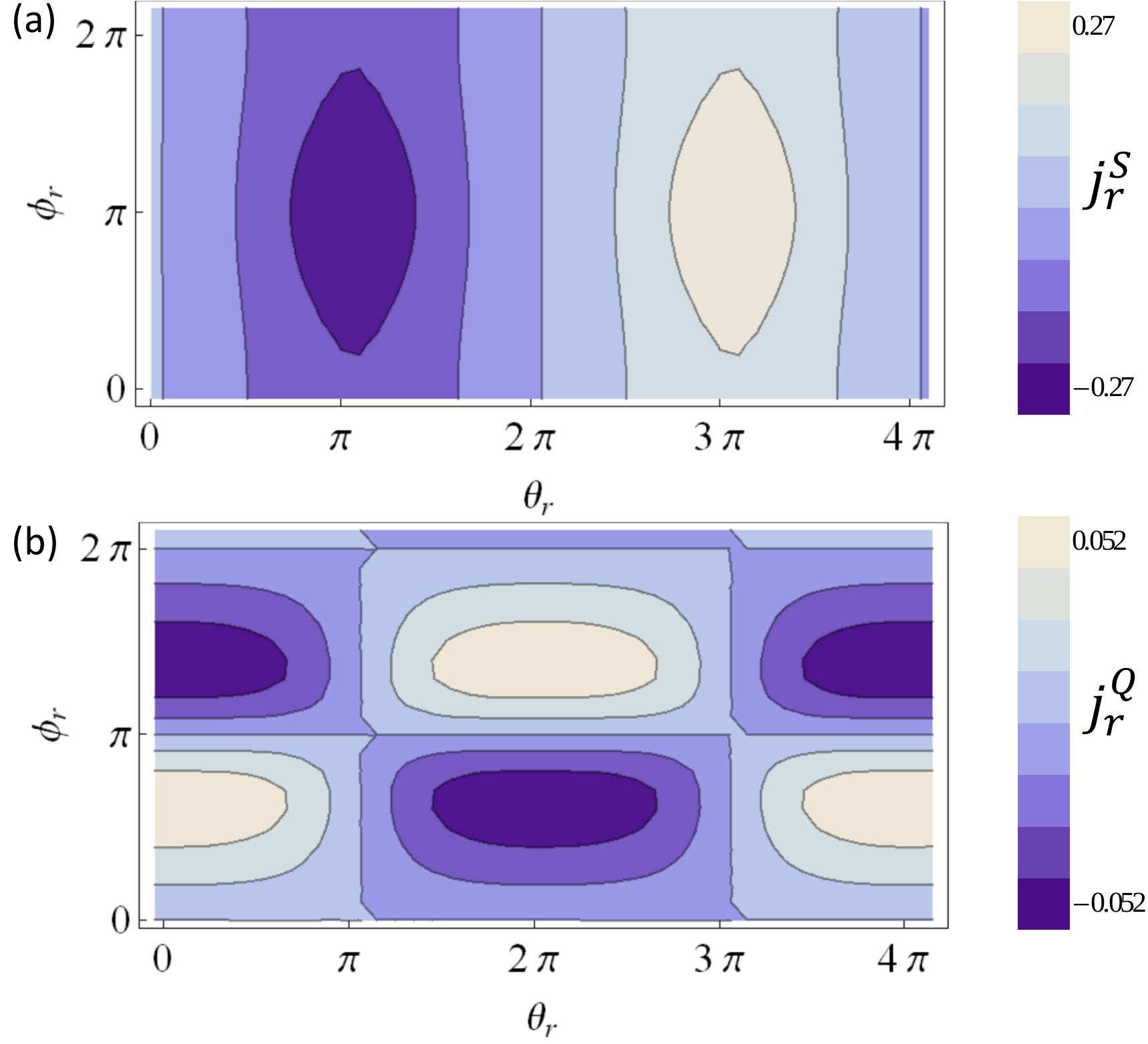} \centering
\caption[fig:ThetaPhi]{Contour plot of (a) spin current $j_{r}^{S}$ and (b)
charge current $j_{r}^{Q}$, both of which are $4\pi$ periodic in $\theta_{r}$
and $2\pi$ periodic in $\phi_{r}$. The other angles are fixed $\phi_{l}=\pi
/2$, $\phi_{m}=\theta_{l/m}=0$. The parameters are the same as
Fig.~\ref{fig:JvsL}. }%
\label{fig:ThetaPhi}%
\end{figure}

Observing the unconventional magneto-Josephson effect and the $4\pi$
periodicity in $\theta_{l/r}$ [see Fig.~\ref{fig:ThetaPhi}(b)] requires
effective control of the magnetic field orientation. In particular, the
orientation change needs to be sufficiently fast so that the Majorana states'
total parity does not change, but still slow on the scale of the inverse bulk
gap to avoid quasiparticle poisoning \cite{SanJose11}. The rate of parity
decay is strongly detail dependent, but we surmise that measurements with
rates faster than $1$ kHz and slower than the minimum gap in the device would
suffice. Conventional magnets may be too unwieldy when made to rapidly turn;
nuclear magnetization, however, could be ideal for this task. Through the
hyperfine coupling, a polarized nuclear spin population could create an
effective Zeeman field in the plane perpendicular to the spin-orbit coupling
direction. For example, large nuclear spin polarization, normal to the
spin-orbit direction, can be induced by optical pumping with circularly
polarized light \cite{Kikkawa00}. An external magnetic field with strength
$b$, applied parallel to the spin-orbit axis, would make the orientation angle
of the hyperfine transverse field wind at a rate $\omega_{L}=\gamma b$, where
$\gamma/2\pi\approx-7.6$ MHz/T for $^{199}$Hg or $\gamma/2\pi\approx
13.5\ $MHz/T for $^{125}$Te nuclei \cite{Willig76}. The hyperfine transverse
field can be rather strong, e.g., $B\sim0.1$ Tesla for $2\%$ nuclear
polarization fraction \cite{Kikkawa00}. It can, moreover, persist for long
times, limited by the inhomogeneous nuclear transverse spin lifetime
$T_{2}^{\ast}\sim100$ $\mu$s, which already suffices for hundreds of
precession periods for $b\sim0.1$ Tesla. The transverse spin lifetime can be
further extended using spin echo techniques.

With a rotating transverse magnetic field, we can observe the
magneto-Josephson effect in several ways. A constantly winding orientation in
the left domain, $\theta_{r}\left(  t\right)  =\omega_{L}t$ [while fixing
$\theta_{l/m}\left(  t\right)  =0$, as illustrated in Fig.~\ref{fig:JvsL}(a)],
produces an oscillatory component of the charge current with amplitude
$j_{\omega_{L}/2}^{Q}=j_{M}^{Q}+j_{Z}^{Q}=\frac{2e}{\hbar}\frac{\partial
E_{0}}{\partial\phi}e^{-L/\lambda_{m-}}$ at \textit{half} the frequency,
$\omega_{L}/2$. In TI edges, we can also use resonant properties to probe the
orientation-frequency halving. A DC voltage $V$ applied to the right
superconducting lead, for instance, induces a winding of the superconducting
angles, $\phi_{r}\left(  t\right)  =2eVt/\hbar$ and $\phi_{l/m}\left(
t\right)  =0$. When the magnetic orientation also winds with angular velocity
$\Omega_{L}$, interference between the two oscillation would yield a DC
current from the right superconducting lead, when $\omega_{L}=2\Omega_{V}$
(neglecting high-order resonances). The amplitude of the dc current is
expected to be:
\begin{equation}
j_{\omega_{L}=2\Omega_{V}}^{Q,DC}\approx\frac{1}{2\pi}\int_{0}^{2\pi}%
j_{\omega_{L}/2}^{Q}\left[  \phi_{r}\right]  \cos\phi_{r}d\phi_{r}.
\end{equation}
Alternatively, one can apply an AC voltage to the right superconducting lead
such that $\phi_{r}\propto\sin\omega t$, while all other superconducting
angles are held fixed. Interference effects now produce Shapiro-step-like
resonant features which emerge only when%
\begin{equation}
\omega_{L}=2n\omega
\end{equation}
for even integer $2n$ (neglecting higher order corrections to the $\theta$ dependence).

The Majorana-mediated spin currents with $4\pi$ phase periodicity are harder
to measure. A possible route for such measurements is to use a magnetic
nanoparticle as the magnetic field source on one of the side domains. The
torques on the nanoparticle could be probed from the shift in the
ferromagnetic resonance (FMR) frequency. The FMR frequency is typically
$f_{0}\sim10$ GHz. The FMR linewidth, dictated by the Gilbert damping
coefficient $\alpha$, is of order $\alpha f_{0}=0.01f_{0}$ in bulk
ferromagnets, but is probably much smaller in nanoparticles \cite{Cehovin03}.
A rough estimate of the maximum Majorana-related spin-current (or torque),
$j^{S}$, yields $j^{S}\sim\hbar\cdot10$ GHz. This produces a frequency shift
around $j^{S}/m_{total}$, which is inversely proportional to the total angular
momentum of the FM grain $m_{total}$ \cite{Brataas12}. This shift must
dominate the FMR linewidth, $j^{S}/m_{total}>f_{0}\alpha$. The nanograin must,
therefore, be sufficiently small such that $m_{total}/\hbar<\alpha^{-1}%
\sim100$, e.g. have a radius of around $10$ nm, and still provide a sufficient
Zeeman field for the domain it is on.


Measuring the effect of the relative field orientation on the spin and charge
currents can be complicated by the presence of conventional Josephson effects
arising from the continuum states. Indeed, the bulk energy associated with the
continuum states also has dependence on magnetic field orientations and
superconducting phases that are interesting in their own right, and of similar
magnitude to the Majorana related effects. Nonetheless, all these dependencies
are $2\pi$ periodic, as we have confirmed numerically. Hence, the measurement
schemes proposed above will be insensitive to them.

In conclusion, we explored consequences of a magnetism-superconductivity
duality of TI edge states, emphasizing Josephson effects. Most prominently,
the duality implies that spin and charge Josephson currents in TI edges
exhibit a $4\pi$ periodic dependence on the orientation difference of the
magnetic field. These remarkable effects are a direct consequence of the
Majorana states and we make several proposals how to detect them
experimentally. The duality is only approximate in spin-orbit-coupled quantum
wires but analogous effects also occur in this system. In addition to the
Josephson effects, the duality has further interesting implications. For
instance, it implies that the transition between topological and trivial
phases can be tuned using a magnetic gradient, which is the dual of the
superconducting phase gradient \cite{Romito12}.


\paragraph{Note added:}

As we are completing the manuscript we became aware of overlap work by Qinglei
Meng, Vasudha Shivamoggi, Taylor Hughes, Matthew Gilbert, and Smitha
Vishveshwara \cite{Meng12}.


It is a pleasure to thank M. P. A. Fisher, L. Glazman, A. Haim, B. Halperin,
A. Kitaev, L. Kouwenhoven, C. Marcus, J. Meyer, Y. Most, F. Pientka, J.
Preskill, X.L. Qi, K. Shtengel, and A. Stern for useful discussions, and the
Aspen Center for Physics for hospitality. We are also grateful for support
from the NSF through grant DMR-1055522, BSF, SPP1285 (DFG), NBRPC (973
program) 2011CBA00300, the Alfred P. Sloan Foundation, the Packard Foundation,
the Humboldt Foundation, the Minerva Foundation, the Sherman Fairchild
Foundation, the Lee A. DuBridge Foundation, the Moore-Foundation funded CEQS,
and the Institute for Quantum Information and Matter (IQIM), an NSF Physics
Frontiers Center with support of the Gordon and Betty Moore Foundation.



\pagebreak\newpage\appendix

\begin{widetext}
\section{Supplementary materials}
We study the linearized 1D system%
\begin{equation}
H\left(  \mu,\Delta,\phi;b,B,\theta\right)  =p\tau^{z}\sigma^{z}-\mu\tau
^{z}+\Delta\left(  \tau^{x}\cos\phi-\tau^{y}\sin\phi\right)  -b\sigma
^{z}+B\left(  \sigma^{x}\cos\theta-\sigma^{y}\sin\theta\right)
\end{equation}
with six control parameters: $\mu$ for the chemical potential, $\Delta$ for
the pairing energy, $\phi$ for the superconducting phase, $-b$ for the
longitudinal magnetic field, $B$ for the transversal magnetic field, and
$\theta$ for angle of the transversal magnetic field. In this form, the
duality between $\left(  \Delta,\mu\right)  $ and $\left(  B,b\right)  $ is
more obvious. Without loss of generality, we assume that all the control
parameters $\left(  \mu,\Delta,b,B\right)  $ are all positive.
\subsection{Phase Diagram}
We compute the determinant%
\begin{equation}
\det H=\left[  p^{2}+\left(  \sqrt{B^{2}-\mu^{2}}+\sqrt{\Delta^{2}-b^{2}%
}\right)  ^{2}\right]  \left[  p^{2}+\left(  \sqrt{B^{2}-\mu^{2}}-\sqrt
{\Delta^{2}-b^{2}}\right)  ^{2}\right]  .
\end{equation}
The energy gap will be closed if there exist some real solutions of $p$ to
satisfy $\det H=0$.
\begin{enumerate}
\item When $B^{2}\leq\mu^{2}$ and $\Delta^{2}\leq b^{2}$, the system is in a
\textbf{gapless-phase}, because there are real solutions $p=\pm\left(
\sqrt{-B^{2}+\mu^{2}}+\sqrt{-\Delta^{2}+b^{2}}\right)  $ or $p=\pm\left(
\sqrt{-B^{2}+\mu^{2}}-\sqrt{-\Delta^{2}+b^{2}}\right)  $ to fulfill the
requirement of $\det H=0$.
\item When $B^{2}>\mu^{2}$ or $\Delta^{2}>b^{2}$, the system is always gapped,
because there are no real solutions of $p$ to satisfy $\det H=0$.
\begin{enumerate}
\item For $\Delta^{2}-b^{2}>\max\left[  B^{2}-\mu^{2},0\right]  $, the system
is in a superconducting gapped phase ($\Delta$\textbf{-phase}).
\item For $B^{2}-\mu^{2}>\max\left[  \Delta^{2}-b^{2},0\right]  $, the system
is in a magnetic gapped phase ($B$\textbf{-phase}).
\item There is a quantum phase transition at $\Delta^{2}-b^{2}=B^{2}-\mu^{2}$,
which connects the $\Delta$-phase and the $B$-phase.
\end{enumerate}
\end{enumerate}
Therefore, we obtain the phase diagram in Fig.~\ref{fig:PhaseDiagram}(a).
\subsection{1D System Consisting of Different Regions}
We are interested in the case that the 1D system consists of three regions of
different control parameters. Specifically%
\begin{equation}
\chi=\left\{
\begin{tabular}
[c]{ll}%
$\chi_{l}$ & for $x\in\left(  -\infty,0\right)  $\\
$\chi_{m}$ & for $x\in\left(  0,L\right)  $\\
$\chi_{r}$ & for $x\in\left(  L,+\infty\right)  $%
\end{tabular}
\right.
\end{equation}
with $\chi$ representing the six control parameters. The system Hamiltonian is%
\begin{equation}
H=\left\{
\begin{tabular}
[c]{ll}%
$H_{l}$ & for $x\in\left(  -\infty,0\right)  $\\
$H_{m}$ & for $x\in\left(  0,L\right)  $\\
$H_{r}$ & for $x\in\left(  L,+\infty\right)  $%
\end{tabular}
\right.
\end{equation}
with $H_{f}\equiv H\left(  \mu_{f},\Delta_{f},\phi_{f};b_{f},B_{f},\theta
_{f}\right)  $. We are interested in the $B-\Delta-B$ configuration, with
$B_{l}^{2}-\mu_{l}^{2}>\max\left[  \Delta_{l}^{2}-b_{l}^{2},0\right]  $,
$\Delta_{m}^{2}-b_{m}^{2}>\max\left[  B_{m}^{2}-\mu_{m}^{2},0\right]  $, and
$B_{r}^{2}-\mu_{r}^{2}>\max\left[  \Delta_{r}^{2}-b_{r}^{2},0\right]  $.
\subsection{Perturbative Formulism for the Coupling Energy}
Let's first consider the individual Majoranas. The left Majorana $\left\vert
L\right\rangle $ is at $x=0$ associated with the $l-m$ boundary. We may
introduce the Hamiltonian $H_{L}=\left\{
\begin{tabular}
[c]{ll}%
$H_{l}$ & for $x\in\left(  -\infty,0\right)  $\\
$H_{m}$ & for $x\in\left(  0,\infty\right)  $%
\end{tabular}
\ \right.  $ that supports the zero energy Majorana mode $\left\vert
L\right\rangle $, with $H_{L}\left\vert L\right\rangle =0.$Similarly, the
right Majorana $\left\vert R\right\rangle $ is at $x=L$ associated with the
$m-r$ boundary. We can also introduce $H_{R}=\left\{
\begin{tabular}
[c]{ll}%
$H_{m}$ & for $x\in\left(  -\infty,L\right)  $\\
$H_{r}$ & for $x\in\left(  L,+\infty\right)  $%
\end{tabular}
\ \right.  $ that supports zero-energy Majorana mode $\left\vert
R\right\rangle $, with $H_{R}\left\vert R\right\rangle =0.$ We can can
perturbatively compute the coupling energy between $\left\vert L\right\rangle
$ and $\left\vert R\right\rangle $ by the formula:
\begin{equation}
{\mathcal{H}}_{LR}\approx M^{-1/2}hM^{-1/2}%
\end{equation}
with $M$ being the overlap matrix between the (not necessarily normalized)
Majorana states, and $h$ being:
\begin{equation}
h=\left(
\begin{array}
[c]{cc}%
0 & \left\langle L\right\vert \Delta V\left\vert R\right\rangle \\
\left\langle R\right\vert \Delta V\left\vert L\right\rangle  & 0
\end{array}
\right)
\end{equation}
with:
\begin{equation}
\Delta V=H-H_{L}=\left(  H_{r}-H_{m}\right)  \eta\left(  x-L\right)  .
\end{equation}
Therefore, the coupling Hamiltonian is approximately ${\mathcal{H}}%
_{LR}\approx\left(
\begin{array}
[c]{cc}%
0 & E\\
E^{\ast} & 0
\end{array}
\right)  $, with
\begin{equation}
E\approx\frac{\left\langle L\right\vert \Delta V_{L}\left\vert R\right\rangle
}{\sqrt{\left\langle L|L\right\rangle \left\langle R|R\right\rangle }}.
\end{equation}
\subsection{Wavefunction of Individual Majoranas}
We can rewrite the Hamiltonian as
\begin{align}
H_{L}  &  =H_{l}\eta\left(  -x\right)  +H_{m}\eta\left(  x\right) \\
&  =\left\{
\begin{tabular}
[c]{ll}%
$U_{l}\cdot V\cdot\left(  p-K_{l}\right)  \cdot V^{\dag}\cdot U_{l}^{\dag}%
\tau^{z}\sigma^{z}$ & for $x<0$\\
$U_{m}\cdot V\cdot\left(  p-K_{m}\right)  \cdot V^{\dag}\cdot U_{m}^{\dag}%
\tau^{z}\sigma^{z}$ & for $x>0$%
\end{tabular}
\ \ \ \right.  .
\end{align}
where the unitary transformations are%
\begin{equation}
V=e^{-i\frac{\pi}{4}\tau^{z}\sigma^{z}}%
\end{equation}%
\begin{equation}
U=e^{i\frac{\phi}{2}\tau^{z}}\otimes e^{i\frac{\theta}{2}\sigma^{z}}%
\equiv:U_{\phi}\otimes U_{\theta}%
\end{equation}
and the non-Hermitian matrix is%
\begin{equation}
K=\left(  b\tau^{z}+i\Delta\tau^{x}\right)  +\left(  \mu\sigma^{z}%
+iB\sigma^{x}\right)
\end{equation}
with sub-index $f=l,m,r$ not explicitly written for simplicity. Without loss
of generality, we can fix
\begin{equation}
\phi_{m}=\theta_{m}=0
\end{equation}
and $U_{m}=I$. For our notational convenience, we also introduce $\tilde
{b}\equiv\cos^{-1}\frac{b}{\Delta}$ and $\tilde{\mu}\equiv\cos^{-1}\frac{\mu
}{B}$. (Let's assume $\Delta^{2}>b^{2}$ and $B^{2}>\mu^{2}$ for notational
simplicity. Later we will show that this constraint can be relaxed.) The
eigensystem of $K$ is%
\begin{equation}
K\cdot\left(  v_{\tilde{b}}^{s_{1}}\otimes v_{\tilde{\mu}}^{s_{2}}\right)
=\Lambda^{s_{1},s_{2}}\left(  v_{\tilde{b}}^{s_{1}}\otimes v_{\tilde{\mu}%
}^{s_{2}}\right)
\end{equation}
with sub-eigenvectors%
\begin{equation}
\fbox{$v_{\xi}^{s}=\frac{1}{\sqrt{2}}\left(  -ie^{is\xi/2},e^{-is\xi
/2}\right)  ^{T}=v_{s\xi}^{+}$}%
\end{equation}
\newline and eigenvalues%
\begin{equation}
\fbox{$\Lambda^{s_{1},s_{2}}=\Delta~\lambda_{\tilde{b}}^{s_{1}}+B~\lambda
_{\tilde{\mu}}^{s_{2}}$}%
\end{equation}
where
\begin{equation}
\lambda_{\xi}^{s}=\lambda_{s\xi}^{+}=i\sin s\xi
\end{equation}
for $s=\pm1$. $\left(  v_{\xi}^{+}\right)  ^{T}\cdot v_{\xi^{\prime}}%
^{+}=-i\sin\frac{\xi+\xi^{\prime}}{2}$. The two-vectors $v_{\xi}^{s}=v_{s\xi
}^{+}$ have the following properties of inner-products:%
\begin{equation}
\left(  v_{\xi}^{s}\right)  ^{T}\cdot v_{\xi}^{s^{\prime}}=-i\sin s\xi
~\delta_{s,s^{\prime}}=\left(
\begin{array}
[c]{cc}%
-i\sin\xi & 0\\
0 & i\sin\xi
\end{array}
\right)
\end{equation}%
\begin{equation}
\left(  v_{\xi}^{s}\right)  ^{\dag}\cdot\sigma^{z}\cdot v_{\xi}^{s^{\prime}%
}=-i\sin s\xi~\delta_{\bar{s},s^{\prime}}=\left(
\begin{array}
[c]{cc}%
0 & -i\sin\xi\\
i\sin\xi & 0
\end{array}
\right)
\end{equation}%
\begin{equation}
\left(  v_{\xi}^{s}\right)  ^{\dag}\cdot v_{\xi}^{s^{\prime}}=\delta
_{s,s^{\prime}}+\cos\xi~\delta_{\bar{s},s^{\prime}}=\left(
\begin{array}
[c]{cc}%
1 & \cos\xi\\
\cos\xi & 1
\end{array}
\right)
\end{equation}
where $\bar{s}:\equiv-s$ for $s=\pm1$. And it transforms under the unitary%
\begin{equation}
\fbox{$U_{\theta}v_{\xi}^{+}=v_{\theta+\xi}^{+}.$}%
\end{equation}
\subsubsection{Left Majorana.}
For the $B-\Delta$ interface at $x=0$, the localized zero-energy eigenstate is%
\begin{equation}
\left\vert L\right\rangle =\left\{
\begin{tabular}
[c]{ll}%
$V\cdot U_{l}\cdot\tau^{z}\sigma^{z}\left\vert \Psi_{\alpha}\right\rangle $ &
for $x<0$\\
$V\cdot U_{m}\cdot\tau^{z}\sigma^{z}\left\vert \Psi_{\beta}\right\rangle $ &
for $x>0$%
\end{tabular}
\ \ \right.  ,
\end{equation}
with%
\begin{align}
\Psi_{\alpha}\left(  x\right)   &  =\sum_{s}v_{\tilde{b}_{l}}^{s}\otimes
v_{\tilde{\mu}_{l}}^{+}~\alpha_{s}e^{-i\Lambda_{l}^{s,+}x}\\
\Psi_{\beta}\left(  x\right)   &  =\sum_{s}v_{\tilde{b}_{m}}^{-}\otimes
v_{\tilde{\mu}_{m}}^{s}~\beta_{s}e^{-i\Lambda_{m}^{-,s}x}.
\end{align}
One can verify%
\begin{equation}
H_{L}\left\vert L\right\rangle =0
\end{equation}
because%
\begin{equation}
\left\{
\begin{tabular}
[c]{ll}%
$\left(  p-K_{l}\right)  \left\vert \Psi_{\alpha}\right\rangle =0$ & for
$x<0$\\
$\left(  p-K_{m}\right)  \left\vert \Psi_{\beta}\right\rangle =0$ & for $x>0$%
\end{tabular}
\ \ \right.  .
\end{equation}
The boundary condition $\left\vert L\left(  x=0^{-}\right)  \right\rangle
=\left\vert L\left(  x=0^{+}\right)  \right\rangle $ requires
\begin{equation}
U_{l}\left\vert \Psi_{\alpha}\left(  x=0^{-}\right)  \right\rangle =\left\vert
\Psi_{\beta}\left(  x=0^{+}\right)  \right\rangle ,
\end{equation}
and hence%
\begin{equation}
\sum_{s}v_{s\tilde{b}_{l}}^{+}~\alpha_{s}=U_{-\phi_{l}}v_{-\tilde{b}_{m}}^{+}%
\end{equation}%
\begin{equation}
\sum_{s}v_{s\tilde{\mu}_{m}}^{+}~\beta_{s}=U_{\theta_{l}}v_{\tilde{\mu}_{l}%
}^{+}%
\end{equation}
which gives us%
\begin{align}
\alpha_{s}  &  =\sin^{-1}s\tilde{b}_{l}\sin\frac{s\tilde{b}_{l}-\left(
\phi_{l}+\tilde{b}_{m}\right)  }{2}\label{eq:alpha}\\
\beta_{s}  &  =\sin^{-1}s\tilde{\mu}_{m}\sin\frac{s\tilde{\mu}_{m}+\left(
\theta_{l}+\tilde{\mu}_{l}\right)  }{2}. \label{eq:beta}%
\end{align}
\subsubsection{Right Majorana.}
Similarly, For the $\Delta-B$ interface at $x=L$, the localized zero-energy
eigenstate is%
\begin{equation}
\left\vert R\right\rangle =\left\{
\begin{tabular}
[c]{ll}%
$V\cdot U_{m}\cdot\tau^{z}\sigma^{z}\left\vert \Psi_{\gamma}\right\rangle $ &
for $x<L$\\
$V\cdot U_{r}\cdot\tau^{z}\sigma^{z}\left\vert \Psi_{\delta}\right\rangle $ &
for $x>R$%
\end{tabular}
\ \ \right.  ,
\end{equation}
with%
\begin{align}
\Psi_{\gamma}\left(  x\right)   &  =\sum_{s}v_{\tilde{b}_{m}}^{+}\otimes
v_{\tilde{\mu}_{m}}^{s}~\gamma_{s}e^{-i\Lambda_{m}^{+,s}\left(  x-L\right)
}\\
\Psi_{\delta}\left(  x\right)   &  =\sum_{s}v_{\tilde{b}_{r}}^{s}\otimes
v_{\tilde{\mu}_{r}}^{-}~\delta_{s}e^{-i\Lambda_{r}^{s,-}\left(  x-L\right)  }.
\end{align}
and
\begin{align}
\gamma_{s}  &  =\sin^{-1}s\tilde{\mu}_{m}\sin\frac{s\tilde{\mu}_{m}+\left(
\theta_{r}-\tilde{\mu}_{r}\right)  }{2}\label{eq:gamma}\\
\delta_{s}  &  =\sin^{-1}s\tilde{b}_{l}\sin\frac{s\tilde{b}_{r}-\left(
\phi_{r}-\tilde{b}_{m}\right)  }{2}. \label{eq:delta}%
\end{align}
\subsection{Normalization of Wavefunctions}
The normalization of wavefunction is%
\begin{align}
M_{l}\left[  \phi_{l}\right]   &  \equiv\left\langle L|L\right\rangle
\nonumber\\
&  =\int_{0}^{\infty}dx\left\langle \Psi_{\beta}\left(  x\right)  |\Psi
_{\beta}\left(  x\right)  \right\rangle +\int_{-\infty}^{0}dx\left\langle
\Psi_{\alpha}\left(  x\right)  |\Psi_{\alpha}\left(  x\right)  \right\rangle
\nonumber\\
&  \approx\frac{\left(  \Delta_{m}^{2}+\mu_{m}^{2}-b_{m}^{2}\right)  }%
{2\sqrt{\Delta_{m}^{2}-b_{m}^{2}}\left(  \Delta_{m}^{2}+\mu_{m}^{2}-B_{m}%
^{2}-b_{m}^{2}\right)  .}\nonumber\\
&  +\frac{\left(  B_{l}^{2}+b_{l}^{2}-\mu_{l}^{2}\right)  +\Delta_{l}\left[
\sqrt{B_{l}^{2}-\mu_{l}^{2}}\sin\left(  \tilde{b}_{m}+\phi_{l}\right)
-b_{l}\cos\left(  \tilde{b}_{m}+\phi_{l}\right)  \right]  }{2\sqrt{B_{l}%
^{2}-\mu_{l}^{2}}\left(  B_{l}^{2}+b_{l}^{2}-\Delta_{l}^{2}-\mu_{l}%
^{2}\right)  } \label{eq:LL}%
\end{align}
\newline Note that the each of the two terms are positive definite, because
$B_{l}^{2}+b_{l}^{2}>\Delta_{l}^{2}+\mu_{l}^{2}$ and $\Delta_{m}^{2}+\mu
_{m}^{2}>B_{m}^{2}+b_{m}^{2}$. By taking $\mu_{f}=0$ (i.e., $\tilde{\mu}%
_{f}=\pi/2$), $b_{l,m,r}=0$ (i.e., $\tilde{b}_{l,m,r}=\pi/2$), we have the
expressions
\begin{equation}
\left\langle L|L\right\rangle _{00}=\frac{\Delta_{m}}{2\left(  \Delta_{m}%
^{2}-B_{m}^{2}\right)  }+\frac{B_{l}+\Delta_{l}\cos\phi_{l}}{2\left(
B_{l}^{2}-\Delta_{l}^{2}\right)  }.
\end{equation}
We can also compute $M_{r}\left[  \phi_{r}\right]  \equiv\left\langle
R|R\right\rangle $, which is very similar to $M_{l}\left[  \phi_{l}\right]  $
with the following replacement
\begin{align}
\tilde{b}_{m}+\phi_{l}  &  \Longrightarrow\tilde{b}_{m}-\phi_{r}\\
\Delta_{l},\mu_{l},B_{l},b_{l}  &  \Longrightarrow\Delta_{r},\mu_{r}%
,B_{r},b_{r}%
\end{align}
\subsection{Cross Coupling $\left\langle L\right\vert \Delta V_{L}\left\vert
R\right\rangle $}
We now compute the cross coupling term $\left\langle L\right\vert \Delta
V_{L}\left\vert R\right\rangle $. First, we can rewrite $\Delta V_{L}$%
\begin{align}
\Delta V_{L} &  =-U_{r}\cdot V\cdot\left(  p-K_{r}\right)  \cdot V^{\dag}\cdot
U_{r}^{\dag}\tau^{z}\sigma^{z}\times\eta\left(  x-L\right)  \nonumber\\
&  +U_{m}\cdot V\cdot\left(  p-K_{m}\right)  \cdot V^{\dag}\cdot U_{m}^{\dag
}\cdot\tau^{z}\sigma^{z}\times\eta\left(  x-L\right)  .
\end{align}
The matrix element
\begin{align}
&  \left\langle L\right\vert \Delta V_{L}\left\vert R\right\rangle \nonumber\\
&  =-\int_{L}^{\infty}dx\left\langle \Psi_{\beta}\left(  x\right)  \right\vert
U_{m}^{\dag}\cdot U_{r}\cdot\tau^{z}\sigma^{z}\cdot K_{r}\left\vert
\Psi_{\delta}\left(  x\right)  \right\rangle +\int_{L}^{\infty}dx\left\langle
\Psi_{\beta}\left(  x\right)  \right\vert \left(  K_{m}\right)  ^{\ast}%
\cdot\tau^{z}\sigma^{z}\cdot U_{m}^{\dag}\cdot U_{r}\left\vert \Psi_{\delta
}\left(  x\right)  \right\rangle \nonumber\\
&  =i\left\langle v_{\tilde{b}_{m}}^{-}\right\vert \tau^{z}\left\vert
v_{\tilde{b}_{m}}^{+}\right\rangle \sum_{s,s^{\prime}}\beta_{s}^{\ast}%
\gamma_{s^{\prime}}e^{\left(  -i\Lambda_{m}^{-,s}\right)  ^{\ast}%
L}\left\langle v_{\tilde{\mu}_{m}}^{s}\right\vert \sigma^{z}\left\vert
v_{\tilde{\mu}_{m}}^{s^{\prime}}\right\rangle \nonumber\\
&  =i\frac{\sin\tilde{b}_{m}}{\sin\tilde{\mu}_{m}}~e^{-\sqrt{\Delta_{m}%
^{2}-b_{m}^{2}}L}\left(
\begin{array}
[c]{c}%
-e^{\sqrt{B_{m}^{2}-\mu_{m}^{2}}L}\sin\frac{\theta_{l}+\tilde{\mu}_{m}%
+\tilde{\mu}_{l}}{2}\sin\frac{\theta_{r}-\tilde{\mu}_{m}-\tilde{\mu}_{r}}{2}\\
+e^{-\sqrt{B_{m}^{2}-\mu_{m}^{2}}L}\sin\frac{\theta_{l}-\tilde{\mu}_{m}%
+\tilde{\mu}_{l}}{2}\sin\frac{\theta_{r}+\tilde{\mu}_{m}-\tilde{\mu}_{r}}{2}%
\end{array}
\right)  \label{eq:LVR}%
\end{align}
By taking $\mu_{l,m,r}=0$ (i.e., $\tilde{\mu}_{l,m,r}=\pi/2$), $b_{l,m,r}=0$
(i.e., $\tilde{b}_{l,m,r}=\pi/2$), we restore the previously obtained familiar
expressions
\begin{equation}
\left\langle L\right\vert \Delta V_{L}\left\vert R\right\rangle _{00}\propto
e^{-\Delta_{m}L}\left(  e^{B_{m}L}\cos\frac{\theta_{l}}{2}\cos\frac{\theta
_{r}}{2}+e^{-B_{m}L}\sin\frac{\theta_{l}}{2}\sin\frac{\theta_{r}}{2}\right)  .
\end{equation}
\subsection{Majorana Coupling Energy}
We can compare the perturbative calculation with the numerical results. For
simplicity, we choose the parameters $\mu_{l/m/r}=0$, $b_{l/m/r}=1/2$,
$\Delta_{m}=2.5$, $\Delta_{l/r}=1$, $B_{l/r}=2$, $B_{m}=1$. The energy from
perturbative calculation is%
\begin{align}
E_{\mathrm{Maj}} &  \approx\frac{\left\langle L\right\vert \Delta
V_{L}\left\vert R\right\rangle }{\sqrt{\left\langle L|L\right\rangle
\left\langle R|R\right\rangle }}\nonumber\\
&  =\frac{1}{\sqrt{M_{l}\left[  \phi_{l}\right]  M_{r}\left[  \phi_{r}\right]
}}\frac{\sin\tilde{b}_{m}}{\sin\tilde{\mu}_{m}}e^{-\sqrt{\Delta_{m}^{2}%
-b_{m}^{2}}L}\left(
\begin{array}
[c]{c}%
-e^{\sqrt{B_{m}^{2}-\mu_{m}^{2}}L}\sin\frac{\theta_{l}+\tilde{\mu}_{m}%
+\tilde{\mu}_{l}}{2}\sin\frac{\theta_{r}-\tilde{\mu}_{m}-\tilde{\mu}_{r}}{2}\\
+e^{-\sqrt{B_{m}^{2}-\mu_{m}^{2}}L}\sin\frac{\theta_{l}-\tilde{\mu}_{m}%
+\tilde{\mu}_{l}}{2}\sin\frac{\theta_{r}+\tilde{\mu}_{m}-\tilde{\mu}_{r}}{2}%
\end{array}
\right)  \label{eq:E}%
\end{align}
For this set of parameters, it will be better to choose $\phi_{l}=\pi/2$,
$\phi_{r}=\pi$, so that $E$ will be most sensitive to the deviation in $\phi$,
which gives the max charge current $I_{Q}\propto\frac{\partial E}{\partial
\phi}$.
\subsection{Analytic Continuation for $\Delta^{2}<b^{2}$ or $B^{2}<\mu^{2}$}
For $\Delta^{2}<b^{2}$ or $B^{2}<\mu^{2}$, we may define the complex number
from the analytic continuation
\begin{equation}
\tilde{b}=\cos^{-1}\frac{b}{\Delta}\equiv-i\cosh^{-1}\frac{b}{\Delta}%
\end{equation}
or
\begin{equation}
\tilde{\mu}=\cos^{-1}\frac{\mu}{B}\equiv-i\cosh^{-1}\frac{\mu}{B}.
\end{equation}
The eigensystem of $K$ has the same form $K\cdot\left(  v_{\tilde{b}}^{s_{1}%
}\otimes v_{\tilde{\mu}}^{s_{2}}\right)  =\Lambda^{s_{1},s_{2}}\left(
v_{\tilde{b}}^{s_{1}}\otimes v_{\tilde{\mu}}^{s_{2}}\right)  $, with
sub-eigenvectors $v_{\xi}^{s}=\frac{1}{\sqrt{2}}\left(  -ie^{is\xi
/2},e^{-is\xi/2}\right)  ^{T}=v_{s\xi}^{+}$, eigenvalues $\Lambda^{s_{1}%
,s_{2}}=\Delta~\lambda_{\tilde{b}}^{s_{1}}+B~\lambda_{\tilde{\mu}}^{s_{2}}$,
and sub-eigenvalues
\begin{equation}
\lambda_{\xi}^{s}=\lambda_{s\xi}^{+}=i\sin s\xi=\sinh si\xi
\end{equation}
for $s=\pm1$. The orthogonality condition remains consistent with the analytic
continuation:%
\begin{equation}
\left(  v_{\xi}^{s}\right)  ^{T}\cdot v_{\xi}^{s^{\prime}}=-i\sin s\xi
~\delta_{s,s^{\prime}}=\left(
\begin{array}
[c]{cc}%
-\sinh i\xi & 0\\
0 & \sinh i\xi
\end{array}
\right)  .
\end{equation}
Hence, the coefficients $\left\{  \alpha_{s},\beta_{s},\gamma_{s},\delta
_{s}\right\}  $ can be obtained by analytic continuation from
Eqs.(\ref{eq:alpha},\ref{eq:beta},\ref{eq:gamma},\ref{eq:delta}). For example,
$\alpha_{s}=\sinh^{-1}si\tilde{b}_{l}\sinh\frac{si\tilde{b}_{l}-\left(
i\phi_{l}+i\tilde{b}_{m}\right)  }{2}$. In order to obtain the the overlap of
wavefunctions, we need to compute $\left(  v_{\xi}^{s}\right)  ^{\dag}$. When
$\cos\xi>1$ ($<1$), $e^{is\xi/2}$ is real (imaginary) and $\left(  v_{\xi}%
^{s}\right)  ^{\dag}=-v_{\xi}^{s}\cdot\sigma^{z}$ ($=i~v_{\xi}^{s}\cdot
\sigma^{x}$). Hence, for $\cos\xi>1$, $\left(  v_{\xi}^{s}\right)  ^{\dag
}\cdot\sigma^{z}\cdot v_{\xi}^{s^{\prime}}=i\sin s\xi~\delta_{s,s^{\prime}}$
and $\left(  v_{\xi}^{s}\right)  ^{\dag}\cdot v_{\xi}^{s^{\prime}}=\cosh
i\xi~\delta_{s,s^{\prime}}+\delta_{\bar{s},s^{\prime}}$. After some careful
calculation, we can verify that the analytic continuation from
Eqs.(\ref{eq:LL},\ref{eq:LVR}) also give the correct results for $\Delta
^{2}<b^{2}$ and/or $B^{2}<\mu^{2}$. Therefore, Eq.(\ref{eq:E}) and its
analytic continuations give the coupling energy between two Majorana bound
states.
\end{widetext}

\end{document}